\documentclass[12pt]{article}
\usepackage{graphicx}
\begin{document}

\begin{center}

{\Large \bf Loop Representation of Wigner's Little Groups}

\vspace{10mm}

Sibel Ba{\c s}kal  \\

Department of Physics, Middle East Technical University, 06800 Ankara, Turkey \\[5ex]

 Young S. Kim \\
Center for Fundamental Physics, University of Maryland College Park,\\ Maryland, MD 20742, USA \\[5ex]

 Marilyn E. Noz \\
Department of Radiology, New York University, New York, NY 10016, USA

\end{center}

\vspace{10mm}

\abstract{Wigner's little groups are the subgroups of the Lorentz group whose
transformations leave the momentum of a given particle invariant.
They thus define the internal space-time symmetries of relativistic
particles. These symmetries take different mathematical forms
for massive and for massless particles. However, it is shown possible to
construct one unified representation using a graphical description. This
graphical approach allows us to describe vividly parity, time reversal, and
charge conjugation of the internal symmetry groups.
As for the language of group theory, the two-by-two representation is used
throughout the paper. While this two-by-two representation is for spin-1/2
particles, it is shown possible to construct the representations for spin-0
particles, spin-1 particles, as well as for higher-spin particles,
for both massive and massless cases. It is shown also that the four-by-four
Dirac matrices constitute a two-by-two representation of Wigner's
little group.}

\newpage

\newpage
\section{Introduction}

In his 1939 paper~\cite{wig39}, Wigner introduced subgroups of the Lorentz
group whose transformations leave the momentum of a given particle invariant.
These subgroups are called Wigner's little groups in the literature
and are known as the symmetry groups for internal space-time structure.

For instance, a massive particle at rest can have spin that can
be rotated in three-dimensional space. The little group in this
case is the three-dimensional rotation group. For a massless particle
moving along the $z$ direction, Wigner noted that rotations around
the $z$ axis do not change the momentum. In addition, he found two more
degrees of freedom, which together with the rotation, constitute a subgroup
locally isomorphic to the two-dimensional Euclidean group.

However, Wigner's 1939 paper did not deal with the following critical
issues.
\begin{itemize}

\item[1.]As for the massive particle, Wigner worked out his little group
   in the Lorentz frame where the particle is at rest with zero momentum,
   resulting in the three-dimensional rotation group.
   He~could have Lorentz-boosted the $O(3)$-like little group to
   make the little group for a moving particle.

\item[2.] While the little group for a massless particle is like $E(2)$,
   it is not difficult to associate the rotational degree of freedom to the
   helicity. However, Wigner did not give physical interpretations
   to the two translation-like degrees of freedom.

\item[3.] While the Lorentz group does not allow mass variations, particles
   with infinite momentum should behave like massless particles.
   The question is whether the Lorentz-boosted $O(3)$-like little
   group becomes the $E(2)$-like little group for
   particles with infinite momentum.

\end{itemize}

\begin{figure}
\centerline{\includegraphics[scale=2.0]{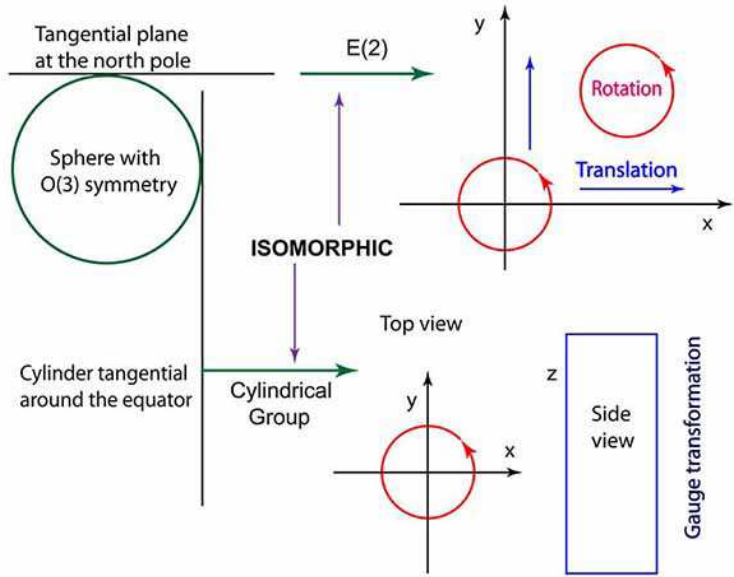}}
\caption{$O(3)$-like and $E(2)$-like internal space-time symmetries of
massive and massless particles.
The sphere corresponds to the $O(3)$-like little group for the massive
particle. There is a plane tangential to the sphere at its north pole,
which is $E(2)$. There is also a cylinder tangent to the sphere at its
equatorial belt. This cylinder gives one helicity and one gauge degree
of freedom. This figure thus gives a unified picture of the little
groups for massive and massless particles~\cite{bkn15}.}\label{iso99}
\end{figure}

 These issues have been properly addressed since
then~\cite{hks83pl,kiwi87jmp,kiwi90jmp,bkn15}. The translation-like degrees
of freedom for massless particles collapse into one gauge degree of freedom,
and the $E(2)$-like little group can be obtained as the infinite-momentum
limit of the $O(3)$-like little group. This history is summarized in
Figure~\ref{iso99}.

%-------------------------------------------------------------------------------
%-------------------------------------------------------------------------------

In this paper, we shall present these developments using a mathematical
language more transparent than those used in earlier papers.
\begin{itemize}
\item[1.] In his original paper~\cite{wig39}, Wigner worked out his little group
 for the massive particle when its momentum is zero. How about moving massive
 particles? In this paper, we start with a moving particle with non-zero
 momentum. We then perform rotations and boosts whose net effect does not
 change the momentum~\cite{kuper76,hk81ajp11,hk88}. This procedure can
 be applied to the massive, massless, and~imaginary-mass cases.

\item[2.] By now, we have a clear understanding of the group $SL(2,c)$ as the
 universal covering group of the Lorentz group. The logic with two-by-two
 matrices is far more transparent than the mathematics based on four-by-four
 matrices. We shall thus use the two-by-two representation of the Lorentz
 group throughout the paper~\cite{bkn15,dir45dub,barg47,naimark54}.
\end{itemize}

The purpose of this paper is to make the physics contained in Wigner's original
paper more transparent.
In Section~\ref{lgroup}, we give the six generators of the Lorentz group. It is
possible to write them in terms of coordinate transformations, four-by-four
matrices, and two-by-two matrices. In Section~\ref{two}, we introduce Wigner's
little groups in terms of two-by-two matrices. In Section~\ref{loops}, it is shown
possible to construct transformation matrices of the little group by performing
rotations and a boost resulting in a non-trivial matrix, which leaves the given
momentum invariant.

 \par

%----------------------------------------------------------------------

%----------------------------------------------------------------------
Since we are more familiar with Dirac matrices than the Lorentz group, it is
shown in Section~\ref{dirac} that Dirac matrices are a representation of the
Lorentz group, and his four-by-four matrices are two-by-two representations
of the two-by-two representation of Wigner's little groups. In Section~\ref{sbt},
we construct spin-0 and spin-1 particles for the SL(2,c) spinors. We
also discuss massless higher spin particles.

\section{Lorentz Group and Its Representations}\label{lgroup}

The group of four-by-four matrices,
which performs Lorentz transformations on the four-dimensional Minkowski space
leaving invariant the quantity $\left(t^2 - z^2 - x^2 - y^2 \right)$, forms the
starting point for the Lorentz group. As there are three rotation and three
boost generators, the Lorentz group is a six-parameter group.

\par
Einstein, by observing that this Lorentz group
also leaves invariant $\left(E, p_z, p_x, p_y \right)$, was able to
derive his Lorentz-covariant energy-momentum relation
commonly known as $E = mc^2$. Thus,
the particle mass is a Lorentz-invariant quantity.

The Lorentz group is generated by the three rotation operators:
\begin{equation}\label{gene01}
 J_{i} = -i\left(x_{j}\frac{\partial}{\partial x_{k}} -
 x_{k}\frac{\partial}{\partial x_{j}}\right) ,
\end{equation}
where $i, j, k = 1, 2, 3$, and three boost operators:
\begin{equation}\label{gene02}
 K_{i} = -i\left(t\frac{\partial}{\partial x_{i}} +
 x_{i}\frac{\partial}{\partial t} \right).
\end{equation}

These generators satisfy the closed set of commutation relations:
\begin{equation}\label{la01}
 \left[J_{i}, J_{j} \right] = i\epsilon_{ijk} J_{k}, \qquad
 \left[J_{i}, K_{j} \right] = i\epsilon_{ijk} K_{k}, \qquad
 \left[K_{i}, K_{j} \right] = - i\epsilon_{ijk} J_{k} ,
\end{equation}
which are known as the Lie algebra for the Lorentz group.

Under the space inversion, $x_{i} \rightarrow -x_{i}$,
or the time reflection, $t \rightarrow -t$, the boost generators $K_{i}$
change sign. However, the Lie algebra remains invariant, which means that
the commutation relations remain invariant under Hermitian conjugation.

In terms of four-by-four matrices applicable to the Minkowskian coordinate
of $(t, z, x, y)$, the~generators can be written as:
\begin{equation}\label{gene01m}
J_{3} = \pmatrix{0 & 0 & 0 & 0 \cr 0 & 0 & 0 & 0 \cr
     0 & 0 & 0 & -i \cr 0 & 0 & i & 0 }, \qquad
K_{3} = \pmatrix{ 0 & i & 0 & 0 \cr i & 0 & 0 & 0 \cr
     0 & 0 & 0 & 0 \cr 0 & 0 & 0 & 0 },
\end{equation}
for rotations around and boosts along the $z$ direction, respectively.
Similar expressions can be written for the $x$ and $y$ directions.
We see here that the rotation generators $J_{i}$ are Hermitian, but
the boost generators $K_{i}$ are anti-Hermitian.
\par

We can also consider the two-by-two matrices:
\begin{equation}\label{gen06}
J_{i} = \frac{1}{2} \sigma_{i}, \quad\mbox{and}\quad
K_{i} = \frac{i}{2} \sigma_{i} ,
\end{equation}
where $\sigma_{i}$ are the Pauli spin matrices. These matrices also
satisfy the commutation relations given in Equation (\ref{la01}).

There are interesting three-parameter subgroups of the Lorentz group.
In 1939~\cite{wig39}, Wigner~considered the subgroups whose transformations
leave the four-momentum of a given particle invariant. First of all,
consider a massive particle at rest. The momentum of this particle
is invariant under rotations in three-dimensional space. What happens
for the massless particle that cannot be brought to a rest frame?
In this paper we shall consider this and other problems using the two-by-two representation
of the Lorentz group.

%---------------------------------------------------------------------

%----------------------------------------------------------------------------------

\section{Two-by-Two Representation of Wigner's Little Groups}\label{two}

 The six generators of Equation (\ref{gen06}) lead to the group of two-by-two unimodular
 matrices of the form:
\begin{equation}\label{alphabeta}
 G = \pmatrix{\alpha & \beta \cr \gamma & \delta} ,
\end{equation}
with $\det(G) = 1$, where the matrix elements are complex numbers. There are
thus six
independent real numbers to accommodate the six generators given in Equation (\ref{gen06}).
The groups of matrices of this form are called $SL(2,c)$ in the literature. Since
the generators $K_{i}$ are not Hermitian, the matrix $G$ is not always unitary.
Its Hermitian conjugate is not necessarily the inverse.

\par
The space-time four-vector can be written as~\cite{bkn15,dir45dub,naimark54}:
\begin{equation}\label{slc01}
\pmatrix{t + z & x - iy \cr x + iy & t - z},
\end{equation}
whose determinant is $t^{2} - z^{2} - x^{2} - z^{2}$, and remains invariant under
the Hermitian transformation:
\begin{equation}\label{slc02}
X' = G~X~G^{\dag} .
\end{equation}

This is thus a Lorentz transformation. This transformation can be explicitly
written as:
\begin{equation}\label{slc04}
\pmatrix{t' + z' & x' - iy' \cr x' + iy' & t' - z'} =
\pmatrix{\alpha & \beta \cr \gamma & \delta}
\pmatrix{t + z & x - iy \cr x + iy & t - z}
\pmatrix{\alpha^* & \gamma^* \cr \beta^* & \delta^*} .
\end{equation}

With these six independent real parameters, it is possible to construct
four-by-four matrices for Lorentz transformations applicable to the four-dimensional
Minkowskian space~\cite{knp86,bkn15}. For the purpose of the present paper,
we need some special cases, and they are given in Table~\ref{tab11}.

%---------------------------------------------------------------------
\begin{table}%[thb]
\caption{Two-by-two and four-by-four representations of the Lorentz group.}\label{tab11}
\vspace{0.5ex}
\begin{center}
\begin{tabular}{lllclc}
\hline
\hline\\[-0.4ex]
\hspace{1mm}& Generators &\hspace{2mm} & Two-by-two &\hspace{2mm} & Four-by-four\\[0.8ex]
\hline\\ [-0.5ex]
\hspace{1mm} &  $J_{3} = \frac{1}{2}\pmatrix{1 & 0 \cr 0  & -1}$
&\hspace{2mm} &  $\pmatrix{\exp{(i\phi/2)} &  0    \cr 0 & \exp{(-i\phi/2)}} $  & \hspace{2mm}&
$\pmatrix{1 & 0 & 0 & 0 \cr 0 & 1 & 0 & 0 \cr 0 & 0 & \cos\phi & -\sin\phi \cr  0 & 0 & \sin\phi & \cos\phi } $
 \\[6.0ex]
\hline\\ [-0.5ex]
\hspace{1mm}&  $K_{3} = \frac{1}{2}\pmatrix{i & 0 \cr 0 & -i}$
&\hspace{2mm} &  $\pmatrix{\exp{(\eta/2)} &  0 \cr 0 & \exp{(-\eta/2)}}$
                & \hspace{2mm} &
$\pmatrix{\cosh\eta & \sinh\eta & 0 & 0 \cr  \sinh\eta & \cosh\eta & 0 & 0 \cr
  0 & 0 & 1 & 0 \cr 0 & 0 & 0 & 1} $
\\[6.0ex]
\hline\\ [-0.5ex]
\hspace{1mm}&  $J_{1} = \frac{1}{2}\pmatrix{0 & 1 \cr 1 & 0}$
&\hspace{2mm} &  $\pmatrix{\cos(\theta/2) &  i\sin(\theta/2)
                  \cr i\sin(\theta/2) & \cos(\theta/2)}$ & \hspace{2mm}&
$\pmatrix{1 & 0 & 0 & 0 \cr 0 & \cos\theta & 0 & \sin\theta  \cr
   0 & 0 & 1 & 0 \cr  0 & -\sin\theta & 0 & \cos\theta }$
\\[6.0ex]
\hline\\ [-0.5ex]
\hspace{1mm}&  $K_{1} = \frac{1}{2}\pmatrix{0 & i \cr i & 0}$
&\hspace{2mm} &  $\pmatrix{\cosh(\lambda/2) &  \sinh(\lambda/2)  \cr \sinh(\lambda/2) & \cosh(\lambda/2)}$ & \hspace{2mm}&
$\pmatrix{\cosh\lambda & 0 & \sinh\lambda & 0 \cr 0 & 1 & 0 & \cr \sinh\lambda & 0 & \cosh\lambda & 0 \cr 0 & 0 & 0 & 1} $
\\[6.0ex]
\hline\\ [-0.5ex]
\hspace{1mm}&  $J_{2} = \frac{1}{2}\pmatrix{0 & -i \cr i & 0}$
&\hspace{2mm} &  $\pmatrix{\cos(\theta/2) &  -\sin(\theta/2)
                  \cr \sin(\theta/2) & \cos(\theta/2)}$ & \hspace{2mm}&
$\pmatrix{1 & 0 & 0 & 0 \cr 0 & \cos\theta & -\sin\theta & 0 \cr
  0 & \sin\theta & \cos\theta & 0 \cr 0 & 0 & 0 & 1} $
\\[6.0ex]
\hline\\ [-0.5ex]
\hspace{1mm}&  $K_{2} = \frac{1}{2}\pmatrix{0 & 1 \cr -1 & 0}$
&\hspace{2mm} &  $\pmatrix{\cosh(\lambda/2) &  -i\sinh(\lambda/2) \cr
                i\sinh(\lambda/2) & \cosh(\lambda/2)}$ & \hspace{2mm}&
$\pmatrix{\cosh\lambda & 0 & 0 & \sinh\lambda  \cr 0 & 1 & 0 & 0 \cr
  0 & 0 & 1 & 0 \cr \sinh\lambda & 0 & 0 & \cosh\lambda} $
\\[6.0ex]
\hline
\hline\\ [-0.5ex]
\end{tabular}
\end{center}
\end{table}
%---------------------------------------------------------------------------------

\section{Two-by-two Representation of Wigner's Little groups}\label{two}

\par
Likewise, the two-by-two matrix for the four-momentum takes the form:
\begin{equation}\label{mom22}
P = \pmatrix{p_0 + p_z & p_x - ip_y \cr p_x + ip_y & p_0 - p_z},
\end{equation}
with $p_0 = \sqrt{m^2 + p_z^2 + p_x^2 + p_2^2}.$
The transformation property of Equation (\ref{slc04}) is applicable also to this
energy-momentum four-vector.

%-------------------------------------------------------------------------------------------

In 1939~\cite{wig39}, Wigner considered the following three four-vectors.
\begin{equation}\label{slc06}
P_{+} = \pmatrix{1 & 0 \cr 0 & 1} , \qquad
P_{0} = \pmatrix{1 & 0 \cr 0 & 0} , \qquad
P_{-} = \pmatrix{1 & 0 \cr 0 & -1} .
\end{equation}
whose determinants are 1, 0, and $-$1, respectively, corresponding to the
four-momenta of massive, massless, and imaginary-mass particles, as shown
in Table~\ref{tab22}.

%---------------------------------------------------------------------------------
\begin{table}
\caption{The Wigner momentum vectors in the two-by-two matrix
representation together with the corresponding transformation matrix.
These four-momentum matrices have determinants that are positive, zero,
and negative for massive, massless, and imaginary-mass particles,
respectively.}\label{tab22}
\vspace{3mm}
\begin{center}
\begin{tabular}{llclc}
\hline \hline\\[0.5ex]
 Particle Mass &{}& Four-Momentum & {} & Transform Matrix \\
\hline\\[-1.0ex]
Massive &{}& $\pmatrix{1 & 0 \cr 0 & 1}$  &{}&
$\pmatrix{\cos(\theta/2) & -\sin(\theta/2)\cr   \sin(\theta/2) & \cos(\theta/2)}$ \\[2.5ex]
 \hline\\[-1.0ex]
Massless &{}&
$\pmatrix{1 & 0 \cr 0 & 0}$ &{}& $\pmatrix{1 & -\gamma \cr 0 & 1}$\\[2.5ex]
  \hline\\[-1.0ex]
Imaginary mass &{}&
$\pmatrix{1 & 0\cr 0 & -1}$
&{}& $\pmatrix{\cosh(\lambda/2) & \sinh(\lambda/2) \cr \sinh(\lambda/2) & \cosh(\lambda/2)}$ \\ [2.5ex]
\hline \hline\\[-0.5ex]
\end{tabular}
\end{center}
\end{table}
%-----------------------------------------------------------------------

He then constructed the subgroups of the Lorentz group whose transformations
leave these four-momenta invariant. These subgroups are called Wigner's little
groups in the literature. Thus, the matrices of these little groups should
satisfy:
\begin{equation}\label{wc00}
 W~P_{i}~W^{\dag} = P_{i} ,
\end{equation}
where $i = +, 0, -$.
Since the momentum of the particle is fixed, these little groups define the
internal space-time symmetries of the particle.
For all three cases, the momentum is invariant under rotations around the $z$
axis, as can be seen from the expression given for the rotation matrix generated
by $J_{3}$ given in Table~\ref{tab11}.

For the first case corresponding to a massive particle at rest, the requirement
of the subgroup is:
\begin{equation}\label{wc01}
 W~P_{+}~W^{\dag} = P_{+} .
\end{equation}

This requirement tells that the subgroup is the rotation subgroup with the
rotation matrix around the $y$ direction:
\begin{equation}\label{wm+}
  R(\theta) = \pmatrix{\cos(\theta/2) & - \sin(\theta/2) \cr
  \sin(\theta/2) & \cos(\theta/2) } .
\end{equation}

For the second case of $P_{0}$, the triangular matrix of the form:
\begin{equation} \label{wm0}
 \Gamma(\xi) = \pmatrix{1 & -\xi \cr 0 & 1} ,
\end{equation}
satisfies the Wigner condition of Equation (\ref{wc00}). If we allow rotations
around the $z$ axis, the expression becomes:
\begin{equation}
\Gamma(\xi,\phi) = \pmatrix{1 & -\xi\exp{(-i\phi)} \cr 0 & 1} .
\end{equation}

This matrix is generated by:
\begin{equation}\label{n12}
N_{1} = J_{2} - K_{1} = \pmatrix{0 & -i \cr 0 & 0} ,
 \quad\mbox{and}\qquad
N_{2} = J_{1} + K_{2} = \pmatrix{0 & 1 \cr 0 & 0} .
\end{equation}

Thus, the little group is generated by $J_{3}$, $N_{1}$, and $N_{2}$. They
satisfy the commutation relations:
\begin{equation}
 \left[N_{1}, N_{2} \right] = 0 , \qquad
 \left[J_{3}, N_{1} \right] = i N_{2}, \qquad
 \left[J_{3}, K_{2} \right] = -i N_{1} . \qquad
\end{equation}

Wigner in 1939~\cite{wig39}
 observed that this set is the same as that of the
two-dimensional Euclidean group with one rotation and two translations.
The physical interpretation of the rotation is easy to understand. It is
the helicity of the massless particle. On the other hand, the physics of
the $N_{1}$ and $N_{2}$ matrices has a stormy history, and the issue was
not completely settled until 1990~\cite{kiwi90jmp}. They generate gauge~
transformations.

\par
For the third case of $P_{-}$, the matrix of the form:
\begin{equation}\label{wm-}
S(\lambda) = \pmatrix{\cosh(\lambda/2) & \sinh(\lambda/2) \cr
 \sinh(\lambda/2) & \cosh(\lambda/2) },
\end{equation}
satisfies the Wigner condition of Equation (\ref{wc00}). This corresponds to
the Lorentz boost along the $x$ direction generated by $K_{1}$ as shown in
Table~\ref{tab11}. Because of the rotation symmetry around the $z$ axis,
the~Wigner condition is satisfied also by the boost along the $y$ axis.
The little group is thus generated by $J_{3}, K_{1}$, and $K_{2}$.
These three generators:
\begin{equation}
 \left[J_{3}, K_{1} \right] = i K_{2}, \qquad
 \left[J_{3}, K_{2} \right] = -i K_{1}, \qquad
 \left[K_{1}, K_{2} \right] = -i J_{3}
\end{equation}
form the little group $O(2,1)$, which is the Lorentz group
applicable to two space-like and one time-like~dimensions.

\par
Of course, we can add rotations around the $z$ axis. Let us Lorentz-boost
these matrices along the $z$ direction with the diagonal matrix:
\begin{equation}
 B(\eta) = \pmatrix{\exp{(\eta/2)} & 0 \cr 0 & \exp{(-\eta/2)} } .
\end{equation}

Then, the matrices of Equations (\ref{wm+}), (\ref{wm0}), and (\ref{wm-}) become:
\begin{eqnarray}
&{}& B(\eta)R(\theta)B(-\eta) = \pmatrix{\cos(\theta/2) &
 - e^{\eta}\sin(\theta/2) \cr
 e^{-\eta}\sin(\theta/2) & \cos(\theta/2) } , \label{wmb+}\\[3ex]
&{}& B(\eta)\Gamma(\xi)B(-\eta) = \pmatrix{1 & -e^{\eta}\xi \cr
 0 & 1}, \label{wmb0} \\[3ex]
&{}& B(\eta)S(-\lambda)B(-\eta)= \pmatrix{\cosh(\lambda/2) &
 - e^{\eta}\sinh(\lambda/2) \cr
 - e^{-\eta}\sinh(\lambda/2) & \cosh(\lambda/2) } , \label{wmb-}
\end{eqnarray}
respectively. We have changed the sign of $\lambda$ for future convenience.

\par
When $\eta$ becomes large, $\theta, \xi$, and $\lambda$ should become small if
the upper-right elements of the these three matrices are to remain finite.
In that case, the diagonal elements become one, and all three matrices become
like the triangular matrix:
\begin{equation}\label{tri01}
 \pmatrix{1 & -\gamma \cr 0 & 1} .
\end{equation}

Here comes the question of whether the matrix of Equation (\ref{wmb-}) can be
continued from Equation~(\ref{wmb+}), via Equation (\ref{wmb0}). For this purpose,
let us write Equation (\ref{wmb+}) as:
\begin{equation}\label{wme+}
\pmatrix{1 - (\gamma\epsilon)^2/2 & - \gamma \cr \gamma\epsilon^2
 & 1 - (\gamma\epsilon)^2/2} ,
\end{equation}
for small $ \theta = 2 \gamma\epsilon $, with $\epsilon = e^{-\eta}$.
For Equation (\ref{wmb-}), we can write:
\begin{equation}\label{wme-}
\pmatrix{1 + (\gamma\epsilon)^2/2 & -\gamma \cr -\gamma\epsilon^2
 & 1 + (\gamma\epsilon)^2/2} ,
\end{equation}
with $\lambda = -2 \gamma\epsilon. $
Both of these expressions become the triangular matrix of Equation (\ref{tri01})
when $\epsilon = 0$.

For small values of $\epsilon$, the diagonal elements change from
$\cos(\theta/2)$ to $\cosh(\lambda/2)$ while $\sin(\theta/2)$ becomes
$-\sinh(\lambda/2)$. Thus, it is possible to continue from Equation (\ref{wmb+})
to Equation (\ref{wmb-}). The~mathematical details of this process have been
discussed in our earlier paper on this subject~\cite{bkn14symm}.

We are then led to the question of whether there is one expression that
will take care of all three cases. We shall discuss this issue in
Section~\ref{loops}.

\section{Loop Representation of Wigner's Little Groups}\label{loops}

It was noted in Section~\ref{two} that matrices of Wigner's little group take
different forms for massive, massless, and imaginary-mass particles. In
this section, we construct one two-by-two matrix that works for all three
different cases.

\par

In his original paper~\cite{wig39}, Wigner constructs those matrices in
specific Lorentz frames. For~instance, for~a moving massive particle with
a non-zero momentum, Wigner brings it to the rest frame and~works out
the $O(3)$ subgroup of the Lorentz group as the little group for this
massive particle. In~order~to complete the little group, we should boost
this $O(3)$ to the frame with the original non-zero momentum~\cite{kiwi90jmp}.

\par
In this section, we construct transformation matrices without changing the
momentum. Let~us assume that the momentum is along the $z$ direction; the
rotation around the $z$ axis leaves the momentum invariant. According to the
Euler decomposition, the rotation around the $y$ axis, in addition, will~
accommodate rotations along all three directions. For this reason, it is
enough to study what happens in transformations within the $xz$
plane~\cite{hks86jmp}.

\par

\begin{figure}
\centerline{\includegraphics[scale=2.6]{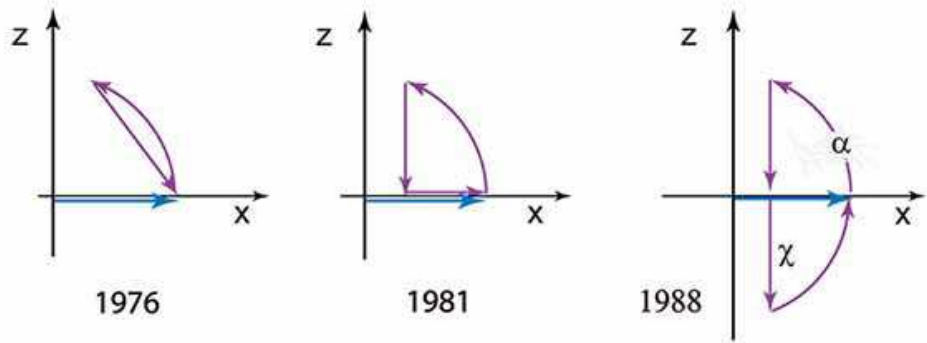}}
\caption{Evolution of the Wigner loop. In 1976~\cite{kuper76}, Kupersztych
considered a rotation followed by a boost whose net result will leave the
momentum invariant. In 1981~\cite{hk81ajp11}, Han and Kim considered the
same problem with simpler forms for boost matrices. In 1988, Han and
Kim~\cite{hk88} constructed the Lorentz kinematics corresponding to the
Bargmann decomposition~\cite{barg47} consisting of one boost matrix
sandwiched by two rotation matrices. In the present case, the two rotation
matrices are identical.}\label{yloop33}
\end{figure}

It was Kupersztych~\cite{kuper76} who showed in 1976 that it is possible
to construct a momentum-preserving transformation by a rotation followed by
a boost as shown in Figure~\ref{yloop33}. In~1981~\cite{hk81ajp11}, Han and
Kim showed that the boost can be decomposed into two components as illustrated
in Figure~\ref{yloop33}. In 1988~\cite{hk88}, Han and Kim showed that the
same purpose can be achieved by one boost preceded and followed by the same
rotation matrix, as shown also in Figure~\ref{yloop33}. We choose to call this
loop the ``D loop'' and write the transformation matrix as:
\begin{equation}\label{d01}
D(\alpha, \chi) = R(\alpha) S(-2\chi) R(\alpha) .
\end{equation}
%----------------------------------------------------------------------
%----------------------------------------------------------------------

The $D$ matrix can now be written as three matrices. This form is known in
the literature as the Bargmann decomposition~\cite{barg47}. This form gives
additional convenience. When we take the inverse or the Hermitian conjugate,
we have to reverse the order of matrices. However, this particular form does
not require re-ordering.

\par
The $D$ matrix of Equation (\ref{d01}) becomes:
\begin{equation} \label{d03}
 D(\alpha, \chi) = \pmatrix{(\cos\alpha)\cosh\chi &
  - \sinh\chi - (\sin\alpha)\cosh\chi
 \cr - \sinh\chi + (\sin\alpha)\cosh\chi & (\cos\alpha)\cosh\chi} .
\end{equation}

If the diagonal element is smaller than one with $((\cos\alpha)\cosh\chi) < 1$,
the off-diagonal elements have opposite signs. Thus, this $D$ matrix can
serve as the Wigner matrix of Equation (\ref{wmb+}) for massive particles. If the
diagonal elements are one, one of the off-diagonal elements vanishes, and this
matrix becomes triangular like Equation (\ref{wmb0}). If the diagonal elements are
greater than one with $((\cos\alpha)\cosh\chi) > 1$, this matrix can become
Equation (\ref{wmb-}). In this way, the matrix of Equation (\ref{d01}) can accommodate
the three different expressions given in Equations (\ref{wmb+})--(\ref{wmb-}).

\subsection{Continuity Problems}

Let us go back to the three separate formulas given in
Equations (\ref{wmb+})--(\ref{wmb-}). If $\eta$ becomes infinity, all three of
them become triangular. For the massive particle, $\tanh\eta$ is the
particle speed, and:
\begin{equation}
  \tanh\eta = \frac{p}{p_{0}},
\end{equation}
where $p$ and $p_{0}$ are the momentum and energy of the particle, respectively.

When the particle is massive with $m^{2} > 0$, the ratio:
\begin{equation}\label{ratio}
 \frac{\mbox{lower-left element}}
 {\mbox{upper-right element}} ,\end{equation}
 is negative and is:
\begin{equation}
-e^{-2\eta} = \frac {1 - \sqrt{1 + m^2/p^2}} {1 + \sqrt{1 + m^2/p^2}} .
\end{equation}

If the mass is imaginary with $m^{2} < 0 $, the ratio is positive and:
\begin{equation}
e^{-2\eta} = \frac {1 - \sqrt{1 + m^2/p^2}} {1 + \sqrt{1 + m^2/p^2}}.
\end{equation}

This ratio is zero for massless particles. This means that when $m^2$ changes
from positive to negative, the ratio changes from $-e^{-2\eta}$
to $e^{-2\eta}$. This transition is continuous, but not analytic. This~
aspect of non-analytic continuity has been discussed in one of our earlier
papers~\cite{bkn14symm}.

The $D$ matrix of Equation (\ref{d03}) combines all three matrices given in
Equations (\ref{wmb+})--(\ref{wmb-}) into one matrix. For this matrix, the ratio of
Equation (\ref{ratio}) becomes:
\begin{equation}
\frac{\tanh\chi - \sin\alpha}{\tanh\chi + \sin\alpha}
= \frac{1- \sqrt{1 + (m/p)^2} }{1 + \sqrt{1 + (m/p^2)}} .
\end{equation}

Thus,
\begin{equation}\label{msq01}
\frac{m^{2}} {p^{2}} = \left(\frac{\sin\alpha}{\tanh\chi}\right)^{2} - 1.
\end{equation}

\begin{figure}
\centerline{\includegraphics[scale=1.2]{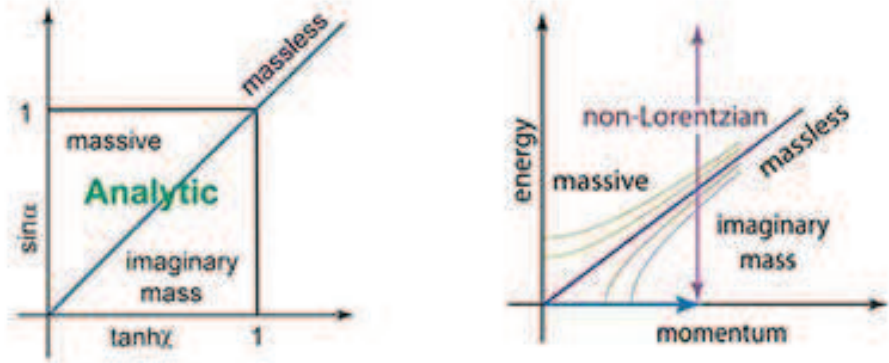}}
\caption{Non-Lorentzian transformations allowing mass variations.
The $D$ matrix of Equation (\ref{d03}) allows us to change the $\chi$ and $\alpha$
analytically within the square region in (\textbf{a}).  These variations allow the
mass variations illustrated in (\textbf{b}), not allowed in Lorentz transformations.
The Lorentz transformations are possible along the hyperbolas given in this
figure. }\label{alphachi}
\end{figure}

For the $D$ loop of Figure~\ref{yloop33}, both $\tanh\chi$ and $\sin\alpha$ range
from 0--1, as illustrated in Figure~\ref{alphachi}.

For small values of the mass for a fixed value of the momentum, this expression
becomes:
\begin{equation}
- \frac{m^{2}}{4p^{2}} .
\end{equation}

Thus, the change from positive values of $m^2$ to negative values is
continuous and analytic. For~massless particles, $m^2$ is zero, while it is
negative for imaginary-mass particles.

We realize that the mass cannot be changed within the frame of the Lorentz
group and that both $\alpha$ and $\eta$ are parameters of the Lorentz
group. On the other hand, their combinations according to the $D$ loop of
Figure~\ref{yloop33} can change the value of $m^{2}$ according to Equation (\ref{msq01})
and Figure~\ref{alphachi}.
%----------------------------------------------------------------------
%----------------------------------------------------------------------

\subsection{Parity, Time Reversal, and Charge Conjugation}

Space inversion leads to the sign change in $\chi$:
 \begin{equation} \label{d11}
 D(\alpha, -\chi) =
 \pmatrix{(\cos\alpha)\cosh\chi & \sinh\chi - (\sin\alpha)\cosh\chi
 \cr \sinh\chi + (\sin\alpha)\cosh\chi & (\cos\alpha)\cosh\chi} ,
 \end{equation}
and time reversal leads to the sign change in both $\alpha$ and $\chi$:
 \begin{equation} \label{d12}
 D(-\alpha, -\chi) =
 \pmatrix{(\cos\alpha)\cosh\chi & \sinh\chi + (\sin\alpha)\cosh\chi
 \cr \sinh\chi - (\sin\alpha)\cosh\chi & (\cos\alpha)\cosh\chi} .
 \end{equation}

If we space-invert this expression, the result is a change only in the direction
of rotation,
 \begin{equation}\label{d14}
 D(-\alpha, \chi) =
 \pmatrix{(\cos\alpha)\cosh\chi & -\sinh\chi + (\sin\alpha)\cosh\chi
 \cr -\sinh\chi - (\sin\alpha)\cosh\chi & (\cos\alpha)\cosh\chi} .
 \end{equation}

\begin{figure}
\centerline{\includegraphics[scale=2.5]{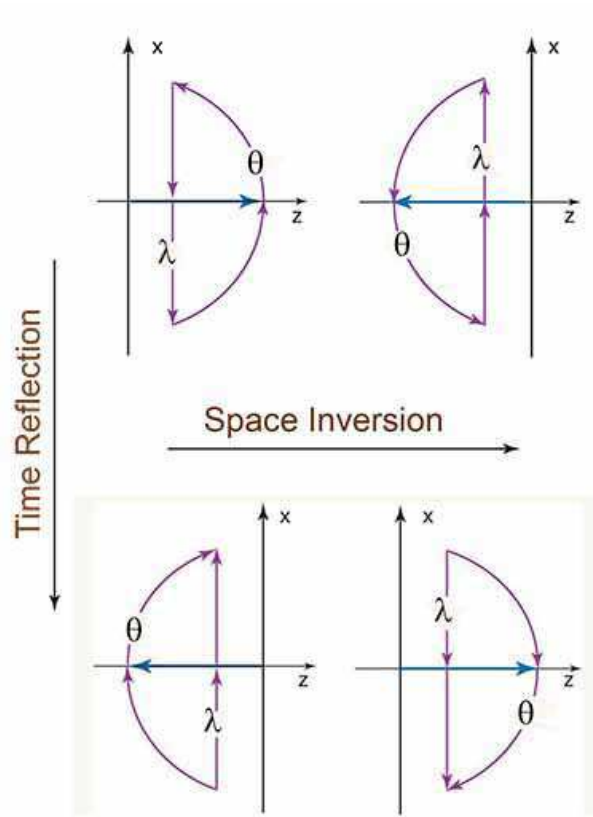}}
\caption{Parity, time reversal, and charge conjugation of Wigner's little
groups in the loop representation.}\label{wloop44}
\end{figure}

The combined transformation of space inversion and time reversal is known
as the ``charge conjugation''. All of these transformations are illustrated in
Figure~\ref{wloop44}.
%----------------------------------------------------------------------
%----------------------------------------------------------------------

Let us go back to the Lie algebra of Equation (\ref{la01}).
This algebra is invariant under Hermitian conjugation. This means that
there is another set of commutation relations,
\begin{equation}\label{la02}
 \left[J_{i}, J_{j} \right] = i\epsilon_{ijk} J_{k}, \qquad
 \left[J_{i}, \dot{K}_{j} \right] = i\epsilon_{ijk} \dot{K}_{k}, \qquad
 \left[\dot{K}_{i}, \dot{K}_{j} \right] = -i\epsilon_{ijk} J_{k} ,
\end{equation}
where $K_{i}$ is replaced with $\dot{K}_{i} = - K_{i}.$ Let us go back to
the expression of Equation (\ref{gene02}). This transition to the dotted
representation is achieved by the space inversion or by the parity operation.

On the other hand, the complex conjugation of the Lie algebra of
Equation (\ref{la01}) leads to:
\begin{equation}\label{la05}
 \left[J_{i}^{*}, J_{j}^{*} \right] = -i\epsilon_{ijk} J_{k}^{*}, \qquad
 \left[J_{i}^{*}, K_{j}^{*} \right] = -i\epsilon_{ijk} K_{k}^{*}, \qquad
 \left[K_{i}^{*}, K_{j}^{*} \right] = i\epsilon_{ijk} J_{k}^{*} .
\end{equation}
It is possible to restore this algebra to that of the original form
of Equation (\ref{la01}) if we replace $J_{i}^{*}$ by $-J_{i}$ and $K_{i}^{*}$
by $ -K_{i}$. This corresponds to the time-reversal process. This
operation is known as the anti-unitary transformation in the
literature~\cite{wig60a,wig60b}.

Since the algebras of Equations (\ref{la01}) and (\ref{la05}) are invariant under
the sign change of $K_{i}$ and $K_{i}^{*}$, respectively, there is another
Lie algebra with $ J_{i}^{*}$ replaced by $-J{i}$ and $K_{i}^{*}$ by
$-\dot{K}_{i}$. This is the parity operation followed by time reversal,
resulting in charge conjugation. With the four-by-four matrices for spin-1
particles, this complex conjugation is trivial, and $J_{i}^{*} = -J_{i}$, as
well as $K_{i}^{*} = - K_{i}.$

On the other hand, for spin $1/2$ particles, we note that:
\begin{eqnarray}
&{}& J_{1}^{*} = J_{1}, \qquad J_{2}^{*} = -J_{2},
\qquad J_{3}^{*} = J_{3}, \nonumber \\[1.0ex]
&{}& K_{1}^{*} = -K_{1}, \qquad K_{2}^{*} = K_{2},
\qquad K_{3}^{*} = - K_{3} .
\end{eqnarray}

 Thus, $J_{i}^{*}$ should be replaced by $\sigma_{2} J_{i}\sigma_{2}$,
 and $K_{i}^{*}$ by -$\sigma_{2} K_{i}\sigma_{2}$.

\section{Dirac Matrices as a Representation of the Little Group}\label{dirac}

The Dirac equation, Dirac matrices, and Dirac spinors constitute the basic
language for spin-1/2 particles in physics. Yet, they are not widely recognized
as the package for Wigner's little group. Yes, the little group is for spins,
so are the Dirac matrices.

Let us write the Dirac equation as:
\begin{equation} \label{de05}
 (p\cdot \gamma - m) \psi(\vec{x}, t) = \lambda \psi(\vec{x}, t) .
\end{equation}

This equation can be explicitly written as:
\begin{equation}\label{de07}
 \left(-i\gamma_{0}\frac{\partial}{\partial t}
 -i\gamma_{1}\frac{\partial}{\partial x}
 -i\gamma_{2}\frac{\partial}{\partial y}
 -i\gamma_{3}\frac{\partial}{\partial z} - m\right) \psi(\vec{x}, t)
 = \lambda \psi(\vec{x}, t) ,
\end{equation}
where:
\begin{equation}
\gamma_{0} = \pmatrix{0 & I \cr I & 0}, \quad
\gamma_{1} = \pmatrix{0 & \sigma_{1} \cr -\sigma_{1} & 0} , \quad
\gamma_{2} = \pmatrix{0 & \sigma_{2} \cr -\sigma_{2} & 0} , \quad
\gamma_{3} = \pmatrix{0 & \sigma_{3} \cr -\sigma_{3} & 0} ,
\end{equation}

where $I$ is the two-by-two unit matrix. We use here the Weyl representation
of the Dirac matrices.

The Dirac spinor has four components. Thus, we write the wave function
for a free particle as:
\begin{equation}\label{de01}
\psi(\vec{x}, t) = U_{\pm}
 \exp{\left[i\left(\vec{p} \cdot \vec{x}- p_{0}t\right)\right]} ,
\end{equation}
with the Dirac spinor:
\begin{equation}\label{de03}
 U_{+} = \pmatrix{ u \cr \dot{u} } , \qquad
 U_{-} = \pmatrix{ v \cr \dot{v} } ,
\end{equation}
where:
\begin{equation}
u = \dot{u} = \pmatrix{1 \cr 0}, \quad\mbox{and}\quad
v = \dot{v} =\pmatrix{0 \cr 1} .
\end{equation}

In Equation (\ref{de01}), the exponential form
$ \exp{\left[i\left(\vec{p} \cdot \vec{x}- p_{0}t\right)\right]}$
defines the particle momentum, and~the column vector $U_{\pm}$ is for the
representation space for Wigner's little group dictating the internal space-time
symmetries of spin-1/2 particles.

\par
In this four-by-four representation, the generators for rotations and boosts
take the form:
\begin{equation}
J_{i} = \frac{1}{2}\pmatrix{\sigma_{i} & 0 \cr 0 & \sigma_{i} },
 \quad\mbox{and}\quad
K_{i} = \frac{i}{2}\pmatrix{\sigma_{i} & 0 \cr 0 & -\sigma_{i} }.
\end{equation}

This means that both dotted and undotted spinor are transformed in the same way
under rotation, while they are boosted in the opposite directions.

\par

When this $\gamma_{0}$ matrix is applied to $U_{\pm}$:
\begin{equation}\label{gamma0}
 \gamma_{0} U_{+} = \pmatrix{0 & I \cr I & 0}
 \pmatrix{ u \cr \dot{u}}
 = \pmatrix{\dot{u} \cr u } , \quad\mbox{and}\quad
 \gamma_{0} U_{-} = \pmatrix{0 & I \cr I & 0}
 \pmatrix{ v \cr \dot{v}}
 = \pmatrix{\dot{v} \cr v } .
\end{equation}

Thus, the $\gamma_{0}$ matrix interchanges the dotted and undotted spinors.
\par
The four-by-four matrix for the rotation around the $y$ axis is:
\begin{equation}
R_{44}(\theta) = \pmatrix{R(\theta) & 0 \cr 0 & R(\theta)} ,
\end{equation}
while the matrix for the boost along the $z$ direction is:
\begin{equation}
B_{44}(\eta) = \pmatrix{B(\eta) & 0 \cr 0 & B(-\eta)} ,
\end{equation}
with:
\begin{equation}
B(\pm \eta) = \pmatrix{e^{{\pm}\eta/2} & 0 \cr 0 & e^{{\mp}\eta/2}} .
\end{equation}

These $\gamma$ matrices satisfy the anticommutation relations:
\begin{equation}
 \left\{\gamma_{\mu} , \gamma_{\nu}\right\} = 2g_{\mu \nu} ,
\end{equation}
 where:
 \begin{eqnarray}
 &{}& g_{00} = 1, \quad g_{11} = g_{22} = g_{22} = -1 , \nonumber \\[1.0ex]
 &{}& g_{\mu \nu} = 0 \quad\mbox{if}\quad \mu \neq \nu .
 \end{eqnarray}

Let us consider space inversion with the exponential form changing to
$\exp{\left[i\left(-\vec{p} \cdot \vec{x}- p_{0}t\right)\right]} $.
For~this purpose, we can change the sign of $x$ in the Dirac equation of
Equation (\ref{de07}). It then becomes:
\begin{equation}\label{de11}
 \left(-i\gamma_{0}\frac{\partial}{\partial t}
 +i\gamma_{1}\frac{\partial}{\partial x}
 +i\gamma_{2}\frac{\partial}{\partial y}
 +i\gamma_{3}\frac{\partial}{\partial z} - m\right) \psi(-\vec{x}, t)
 = \lambda \psi(-\vec{x}, t) .
\end{equation}

Since $\gamma_{0}\gamma_{i} = - \gamma_{i}\gamma_{0}$ for $i = 1,2,3$,
\begin{equation}\label{de15}
 \left(-i\gamma_{0}\frac{\partial}{\partial t}
 -i\gamma_{1}\frac{\partial}{\partial x}
 -i\gamma_{2}\frac{\partial}{\partial y}
 -i\gamma_{3}\frac{\partial}{\partial z} - m\right) [\gamma_{0}
 \psi(-\vec{x}\cdot\vec{p}, p_{0}t)]
 = \lambda [\gamma_{0} \psi(-\vec{x}\cdot\vec{p}, p_{0}t)].
\end{equation}

This is the Dirac equation for the wave function under the space inversion or
the parity operation. The Dirac spinor $U_{\pm}$ becomes $\gamma_{0}U_{\pm}$,
according to Equation (\ref{gamma0}). This operation is illustrated in
Table~\ref{tab33} and Figure~\ref{wloop44}.

%----------------------------------------------------------------------------------
\begin{table}
\caption{Parity, charge conjugation, and time reversal in the loop representation.}\label{tab33}
 \vspace{3mm}
\begin{center}
\begin{tabular}{cccccc}
\hline \\[-2.5ex] \hline \\[-0.5ex]
 \hspace{3mm} &\hspace{5mm} & Start & \hspace{5mm}& Time Reflection & \\[2ex]
 \hline \\[-1.0ex]
 \hspace{3mm} Start &&
$\matrix{ \mbox{Start with} \\R(\alpha)S(-2\chi)R(\alpha)}$
  &{}&
$\matrix{\mbox{Time Reversal} \\R(-\alpha)S(2\chi)R(-\alpha) }$ &
\hspace{2mm}\\[4.0ex]
\hline\\[-1.0ex]
\hspace{3mm}$\matrix{\mbox{Space} \\\mbox{Inversion} }$ &\hspace{5mm}&
$\matrix{\mbox{Parity}
\\R(\alpha)S(2\chi)R(\alpha) }$ &\hspace{5mm} &
$\matrix{\mbox{Charge Conjugation}
\\R(-\alpha)S(-2\chi)R(-\alpha) }$ & \hspace{2mm} \\[4.0ex]
\hline
\hline\\[-0.5ex]
\end{tabular}
\end{center}
\end{table}
%----------------------------------------------------------------------------------

We are interested in changing the sign of $t$. First, we can change
both space and time variables, and then, we can change the space variable.
We can take the complex conjugate of the equation first. Since
$\gamma_{2}$ is imaginary, while all others are real,
the Dirac equation becomes:
\begin{equation}\label{d05}
 \left(i\gamma_{0}\frac{\partial}{\partial t}
+ i\gamma_{1}\frac{\partial}{\partial x}
 - i\gamma_{2}\frac{\partial}{\partial y}
+ i\gamma_{3}\frac{\partial}{\partial z} - m\right) \psi^{*}(\vec{x}, t)
 = \lambda \psi^{*} (\vec{x}, t) .
\end{equation}

We are now interested in restoring this equation to the original form of
Equation (\ref{de07}). In order to achieve this goal, let us consider
$\left(\gamma_{1}\gamma_{3}\right).$ This form commutes with $\gamma_{0}$
and $\gamma_{2}$ and anti-commutes with $\gamma_{1}$ and $\gamma_{3}$.
Thus,
\begin{equation}\label{de17}
 \left(-i\gamma_{0}\frac{\partial}{\partial t}
 -i\gamma_{1}\frac{\partial}{\partial x}
 -i\gamma_{2}\frac{\partial}{\partial y}
 -i\gamma_{3}\frac{\partial}{\partial z} - m\right)
 \left(\gamma_{1}\gamma_{3}\right) \psi^{*}(\vec{x}, t)
 = \lambda \left(\gamma_{1}\gamma_{3}\right) \psi^{*} (\vec{x}, -t) .
\end{equation}

Furthermore, since:
\begin{equation}
\gamma_{1}\gamma_{3} = \pmatrix{i\sigma_{2} & 0 \cr 0 & i\sigma_{2}},\end{equation}
this four-by-four matrix changes the direction of the spin. Indeed,
this form of time reversal is consistent with Table~\ref{tab33}
and Figure~\ref{wloop44}.

Finally, let us change the signs of both $\vec{x}$ and $t$. For
this purpose, we go back to the complex-conjugated Dirac equation of
Equation (\ref{de05}). Here, $\gamma_{2}$ anti-commutes with all others. Thus,
the~wave function:
\begin{equation}
\gamma_{2}\psi(-\vec{x}\cdot\vec{p}, -p_{0}t) ,
\end{equation}
should satisfy the Dirac equation. This form is known as the
charge-conjugated wave function, and it is also illustrated in
Table~\ref{tab33} and Figure~\ref{wloop44}.

\subsection{Polarization of Massless Neutrinos}\label{zeromass}

For massless neutrinos, the little group consists of rotations around the $z$ axis,
in addition to $N_{i}$ and $\dot{N_{i}}$ applicable to the upper and lower
components of the Dirac spinors. Thus, the four-by-four matrix for these
generators is:
\begin{equation}
  N_{44(i)} = \pmatrix{N_{i} & 0 \cr 0 & \dot{N}_{i}} .
\end{equation}

 The transformation matrix is thus:
\begin{equation}
 D_{44}(\alpha,\beta) =
 \exp{\left(-i\alpha N_{44(1)} - i\beta N_{44(2)} \right) }
 = \pmatrix{D(\alpha, \beta) & 0 \cr 0 & \dot{D}(\alpha,\beta)},
\end{equation}
with:
\begin{eqnarray}
D(\alpha,\beta) = \pmatrix{ 1 & \alpha -i\beta \cr 0 & 1 } , \qquad
\dot{D}(\alpha,\beta) \pmatrix{ 1 & 0 \cr -\alpha -i\beta & 1 }.
\end{eqnarray}

As is illustrated in Figure~\ref{iso99}, the $D$ transformation performs
the gauge transformation on massless photons. Thus, this transformation
allows us to extend the concept of gauge transformations to massless
spin-1/2 particles. With this point in mind, let us see what happens
when this $D$ transformation is applied to the Dirac spinors.
\begin{equation}\label{com8}
D(\alpha,\beta)u = u,
\qquad \dot{D}(\alpha,\beta)\dot{v} = \dot{v} .
\end{equation}

Thus, $u$ and $\dot{v}$ are invariant gauge transformations.

What happens to $v$ and $\dot{u}$?
\begin{equation}\label{com9}
D(\alpha,\beta)v = v + (\alpha-i\beta)u,
\qquad \dot{D}(\alpha,\beta)\dot{u} =
\dot{u} - (\alpha + i\beta)\dot{v} .
\end{equation}

These spinors are not invariant under gauge transformations~\cite{hks82,png86}.

Thus, the Dirac spinor:
\begin{equation}
U_{inv} = \pmatrix{ u \cr \dot{v}},
\end{equation}
is gauge-invariant while the spinor
\begin{equation}
U_{non} = \pmatrix{ v \cr \dot{u}},
\end{equation}
is not. Thus, gauge invariance leads to the polarization of massless
spin-1/2 particles. Indeed, this is what we observe in the real world.

\subsection{Small-Mass Neutrinos} \label{smallm}

Neutrino oscillation experiments presently suggest that neutrinos have a small,
but finite mass~\cite{mohap06}. If neutrinos have mass, there should be a
Lorentz frame in which they can be brought to rest with an $O(3)$-like $SU(2)$
little group for their internal space-time symmetry. However, it is not likely
that at-rest neutrinos will be found anytime soon. In the meantime, we have to work
with the neutrino with a fixed momentum and a small mass~\cite{kmn16adv}. Indeed,
the present loop representation is suitable for this problem.

Since the mass is so small, it is appropriate to approach this small-mass problem
as a departure from the massless case. In Section~\ref{zeromass}, it was noted that
the polarization of massless neutrinos is a consequence of gauge invariance.
Let us start with a left-handed massless neutrino with the spinor:
\begin{equation}\label{spinor51}
 \dot{v} = \pmatrix{0 \cr 1} ,
\end{equation}
 and the gauge transformation applicable to this spinor:
 \begin{equation}\label{gauge51}
 \dot{\Gamma}(\gamma) = \pmatrix{1 & 0 \cr \gamma & 1}.
 \end{equation}

Since:
 \begin{equation}
 \pmatrix{1 & 0 \cr \gamma & 1} \pmatrix{0 \cr 1}
 = \pmatrix{0 \cr 1} ,
 \end{equation}
the spinor of Equation (\ref{spinor51}) is invariant under the gauge transformation
of Equation (\ref{gauge51}).

If the neutrino has a small mass, the transformation matrix is for a rotation.
However, for a small non-zero mass, the deviation from the triangular form
is small. The procedure for deriving the Wigner matrix for this case is
given toward the end of Section~\ref{two}. The matrix in this case is:
\begin{equation}
\dot{D}(\gamma) = \pmatrix{1 - (\gamma\epsilon)^2/2 & - \gamma\epsilon^2
 \cr \gamma & 1 - (\gamma\epsilon)^2/2} ,
\end{equation}
with $\epsilon^2 = m/p$, where $m$ and $p$ are the mass and momentum of the
neutrino, respectively. This matrix becomes the gauge transformation of
Equation (\ref{gauge51}) for $\epsilon = 0$. If this matrix is applied to
the spinor of Equation (\ref{spinor51}), it becomes:
\begin{equation}
 \dot{D(\gamma)} \dot{v} = \pmatrix{ - \gamma\epsilon^2 \cr 1}.
\end{equation}

In this way, the left-handed neutrino gains a right-handed component.
We took into account that $(\gamma \epsilon)^2$ is much smaller than one.

Since massless neutrinos are gauge independent, we cannot measure the
value of $\gamma$. For the small-mass case, we can determine this value
from the measured values of $m/p$ and the density of right-handed neutrinos.

\section{Scalars, Vectors, and Tensors}\label{sbt}

We are quite familiar with the process of constructing three spin-1 states
and one spin-0 state from two spinors. Since each spinor has two states,
there are four states if combined.
\par
In the Lorentz-covariant world, for each spin-1/2 particle, there are two
additional two-component spinors coming from the dotted
representation~\cite{knp86,berest82,gelfand63,wein64a}. There are thus
four states. If two spinors are combined, there are 16 states. In this
section, we show that they can be partitioned into
\begin{itemize}
 \item[1.] scalar with one state,
 \item[2.] pseudo-scalar with one state,
 \item[3.] four-vector with four states,
 \item[4.] axial vector with four states,
 \item[5.] second-rank tensor with six states.
\end{itemize}

 These quantities contain sixteen states. We made an attempt to
construct these quantities in our earlier publication~\cite{bkn15}, but this
earlier version is not complete. There, we did not take into account the
parity operation properly. We thus propose to complete the job in this section.

\par
For particles at rest, it is known that the addition of two one-half spins
result in spin-zero and spin-one states. Hence, we have two different
spinors behaving differently under the Lorentz boost. Around the
$z$ direction, both spinors are transformed by:
\begin{equation}\label{max01}
Z(\phi) = \exp{\left(-i\phi J_3\right)}
= \pmatrix{e^{-i\phi/2} & 0 \cr 0 & e^{i\phi/2}}.
\end{equation}

However, they are boosted by:
\begin{eqnarray}\label{max03}
&{}& B(\eta) = \exp{\left(-i\eta K_3\right)}
= \pmatrix{e^{\eta/2} & 0 \cr 0 & e^{-\eta/2}} , \nonumber\\[2ex]
&{}& \dot{B}(\eta) = \exp{\left(i\eta K_3\right)} ,
= \pmatrix{e^{-\eta/2} & 0 \cr 0 & e^{\eta/2}} ,
\end{eqnarray}
which are applicable to the undotted and dotted spinors, respectively.
These two matrices commute with each other and also with the rotation matrix
$Z(\phi)$ of Equation~(\ref{max01}). Since $K_3$ and $J_3$ commute with each
other, we can work with the matrix $Q(\eta,\phi)$ defined as:
\begin{eqnarray}\label{max05}
&{}& Q(\eta,\phi) = B(\eta)Z(\phi) = \pmatrix{e^{(\eta - i\phi)/2} & 0 \cr
 0 & e^{-(\eta - i\phi)/2}} , \nonumber \\[2ex]
&{}& \dot{Q}(\eta, \phi) = \dot{B}(\eta) \dot{Z}(\phi)
= \pmatrix{e^{-(\eta + i\phi)/2} & 0 \cr
 0 & e^{(\eta + i\phi)/2}} .
\end{eqnarray}

\par
When this combined matrix is applied to the spinors,
\begin{eqnarray}\label{max07}
&{}& Q(\eta,\phi) u = e^{(\eta -i\phi)/2} u , \qquad
Q(\eta,\phi) v = e^{-(\eta -i\phi)/2} v, \nonumber\\[2ex]
&{}& \dot{Q}(\eta, \phi) \dot{u} = e^{-(\eta + i\phi)/2} \dot u , \qquad
 \dot{Q}(\eta,\phi)\dot{v} = e^{(\eta + i\phi)/2} \dot{v}.
\end{eqnarray}

\par
If the particle is at rest, we can explicitly construct the combinations:
\begin{equation}\label{max09}
 uu, \qquad \frac{1}{\sqrt{2}}(uv + vu),
\qquad vv,
\end{equation}
to obtain the spin-1 state and:
\begin{equation} \label{max10}
\frac{1}{\sqrt{2}}(uv - vu) ,
\end{equation}
for the spin-zero state. This results in four bilinear states. In the
$SL(2,c)$ regime, there are two dotted spinors, which result in four more
bilinear states. If we include both dotted and undotted spinors, there~are
sixteen independent bilinear combinations. They are given in Table~\ref{tab77}. This
table also gives the effect of the operation of $Q(\eta, \phi)$.

%---------------------------------------------------------------------------------
\begin{table}
\caption{Sixteen combinations of the $SL(2,c)$ spinors. In the $SU(2)$
regime, there are two spinors leading to four bilinear forms. In the $SL(2,c)$
world, there are two undotted and two dotted spinors. These four-spinors lead
to sixteen independent bilinear combinations.}\label{tab77}
\vspace{3mm}
\begin{center}
\begin{tabular}{ccccc}
 \hline \\[-2.5ex] \hline\\ [-0.7ex]
{}& Spin 1 & \hspace{10mm} & Spin 0 & {}
\\[0.9ex] \hline\\[-0.9ex]
{}& $ uu,\quad \frac{1}{\sqrt{2}}(uv + vu), \quad vv, $ & {} &
$\frac{1}{\sqrt{2}}(uv - vu) $ & {}
\\[0.7ex] \hline\\[-0.7ex]
{}& $ \dot{u}\dot{u},\quad \frac{1}{\sqrt{2}}(\dot{u}\dot{v} + \dot{v}\dot{u}),
 \quad \dot{v}\dot{v}, $ & {} &
$\frac{1}{\sqrt{2}}(\dot{u}\dot{v} - \dot{v}\dot{u}) $ & {}
\\[0.7ex] \hline\\[-0.7ex]
{}& $ u\dot{u},\quad \frac{1}{\sqrt{2}}(u\dot{v} + v\dot{u}),
\quad v\dot{v}, $ & {} &
$\frac{1}{\sqrt{2}}(u\dot{v} - v\dot{u}) $ & {}
\\[0.7ex] \hline\\[-0.7ex]
{}& $\dot{u}u, \quad \frac{1}{\sqrt{2}}(\dot{u}v + \dot{v}u),
\quad \dot{v}v, $ & {} &
$\frac{1}{\sqrt{2}}( \dot{u}v - \dot{v}u) $ & {}
\\[0.7ex] \hline\\[-2.5ex]\hline\\[-0.8ex]
{}& After the operation of $Q(\eta,\phi)$ and $\dot{Q}(\eta,\phi)$
 & \hspace{10mm} & & {}
\\[0.0ex] \hline\\[-0.7ex]
{}& $ e^{-i\phi} e^{\eta} uu,\quad \frac{1}{\sqrt{2}}(uv + vu),
\quad e^{i\phi}e^{-\eta} vv,
$ & {} &
$\frac{1}{\sqrt{2}}(uv - vu) $ & {}
\\[0.7ex] \hline\\[-0.7ex]
{}& $ e^{-i\phi}e^{-\eta} \dot{u}\dot{u},\quad
\frac{1}{\sqrt{2}}(\dot{u}\dot{v} +\dot{v}\dot{u}),
 \quad e^{i\phi}e^{\eta} \dot{v}\dot{v}, $ & {} &
$\frac{1}{\sqrt{2}}(\dot{u}\dot{v} - \dot{v}\dot{u}) $ & {}
\\[0.7ex] \hline\\[-0.7ex]
{}& $ e^{-i\phi} u\dot{u},\quad
\frac{1}{\sqrt{2}}(e^{\eta} u\dot{v} + e^{-\eta} v\dot{u}),
\quad e^{i\phi} v\dot{v}, $ & {} &
$\frac{1}{\sqrt{2}}(e^{\eta} u\dot{v} - e^{-\eta} v\dot{u}) $ & {}
\\[0.7ex] \hline\\[-0.7ex]
{}& $e^{-i\phi} \dot{u}u, \quad \frac{1}{\sqrt{2}}(\dot{u}v + \dot{v}u),
\quad e^{i\phi} \dot{v}v, $ & {} &
$\frac{1}{\sqrt{2}}(e^{-\eta} \dot{u}v - e^{\eta} \dot{v}u) $ & {}
\\[0.7ex] \hline \\[-2.5ex] \hline
\end{tabular}
\end{center}
\end{table}
%-----------------------------------------------------------------------

\par
Among the bilinear combinations given in Table~\ref{tab77}, the following two
equations are invariant under rotations and also under boosts:
\begin{equation}\label{max11}
S = \frac{1}{\sqrt{2}}(uv - vu), \quad\mbox{and}\quad
\dot{S} = - \frac{1}{\sqrt{2}}(\dot{u}\dot{v} - \dot{v}\dot{u}) .
\end{equation}

They are thus scalars in the Lorentz-covariant world. Are they the same or
different? Let us consider the following combinations
\par
\begin{equation}\label{max12}
S_{+} = \frac{1}{\sqrt{2}}\left(S + \dot{S}\right), \quad\mbox{and}\quad
S_{-} = \frac{1}{\sqrt{2}}\left(S - \dot{S}\right) .
\end{equation}

Under the dot conjugation, $S_{+}$ remains invariant, but $S_{-}$ changes
sign. The boost is performed in the opposite
direction and therefore is the operation of space inversion.
Thus, $S_{+}$ is a scalar, while $S_{-}$ is called a pseudo-scalar.

%-----------------------------------------------------------------------

\subsection{Four-Vectors} \label{4vec}

Let us go back to Equation (\ref{max09}) and make a dot-conjugation on one of the spinors.
\begin{eqnarray}\label{max13}
&{}& u\dot{u},
\qquad \frac{1}{\sqrt{2}}(u\dot{v} + v \dot{u}), \qquad v\dot{v} ,
\qquad \frac{1}{\sqrt{2}}(u\dot{v} - v \dot{u}),
 \nonumber \\[2ex]
&{}& \dot{u}u,
\qquad \frac{1}{\sqrt{2}}(\dot{u}v + \dot{v}u), \qquad \dot{v}v ,
\qquad \frac{1}{\sqrt{2}}(\dot{u}v - \dot{v}u).
\end{eqnarray}

We can make symmetric combinations under dot conjugation, which lead to:
\begin{eqnarray}\label{max15}
&{}& \frac{1}{\sqrt{2}}\left(u\dot{u} +\dot{u}u\right),
\quad \frac{1}{2}[(u\dot{v} + v \dot{u}) + (\dot{u}v + \dot{v}u)],
 \quad \frac{1}{\sqrt{2}} ( v\dot{v}+ \dot{v}v) ,
 \quad\mbox{for spin 1} , \nonumber \\[2ex]
&{}& \frac{1}{2}[(u\dot{v} - v \dot{u}) + (\dot{u}v - \dot{v}u)],
 \quad\mbox{for spin 0} ,
\end{eqnarray}
and anti-symmetric combinations, which lead to:
\begin{eqnarray}\label{max16}
&{}& \frac{1}{\sqrt{2}}\left(u\dot{u} - \dot{u}u\right),
\quad \frac{1}{2}[(u\dot{v} + v \dot{u}) - (\dot{u}v + \dot{v}u)],
 \quad \frac{1}{\sqrt{2}}(v\dot{v} - \dot{v}v) ,
 \quad\mbox{for spin 1}, \nonumber \\[2ex]
&{}& \frac{1}{2}[(u\dot{v} - v \dot{u}) - (\dot{u}v - \dot{v}u)],
 \quad\mbox{for spin 0} .
\end{eqnarray}

\par

Let us rewrite the expression for the space-time four-vector given in
Equation (\ref{slc01}) as:
\begin{equation}\label{max21}
\pmatrix{t + z & x - iy \cr x + iy & t - z},
\end{equation}
which, under the parity operation, becomes
\begin{equation}\label{max22}
\pmatrix{t - z & -x + iy \cr - x - iy & t + z} .
\end{equation}

If the expression of Equation (\ref{max21}) is for an axial vector, the parity
operation leads to:
\begin{equation}\label{max23}
\pmatrix{- t + z & x - iy \cr x + iy & - t - z},
\end{equation}
where only the sign of $t$ is changed. The off-diagonal elements remain invariant,
while the diagonal elements are interchanged with sign changes.

We note here that the parity operation corresponds to dot conjugation. Then,
from the expressions given in Equations (\ref{max15}) and (\ref{max16}), it is possible
to construct the four-vector as:
\begin{equation}
V = \pmatrix{u\dot{v} - \dot{v}u & v\dot{v} - \dot{v}v \cr
 u\dot{u} - \dot{u}u & \dot{u}v - v\dot{u}} ,
\end{equation}
where the off-diagonal elements change their signs under the dot conjugation, while
the diagonal elements are interchanged.

The axial vector can be written as:
\begin{equation}
A = \pmatrix{u\dot{v} + \dot{v}u & v\dot{v} + \dot{v}v \cr
 u\dot{u} + \dot{u}u & -\dot{u}v - v\dot{u}} .
\end{equation}

Here, the off-diagonal elements do not change their signs under dot conjugation,
and the diagonal elements become interchanged with a sign change. This matrix
thus represents an axial vector.

\subsection{Second-Rank Tensor}\label{tensor}

There are also bilinear spinors, which are both dotted or both undotted. We
are interested in two sets of three quantities satisfying the $O(3)$ symmetry.
They should therefore transform like:
\begin{equation}
(x + iy)/\sqrt{2}, \qquad (x - iy)/\sqrt{2}, \qquad z ,
\end{equation}
which are like:
\begin{equation}
 uu, \qquad vv, \quad (uv + vu)/\sqrt{2} ,
\end{equation}
respectively, in the $O(3)$ regime. Since the dot conjugation is
the parity operation, they are like:
\begin{equation}
-\dot{u}\dot{u}, \qquad -\dot{v}\dot{v}, \qquad
-(\dot{u}\dot{v} + \dot{v}\dot{u})/\sqrt{2} .
\end{equation}

In other words,
\begin{equation}
(uu{\dot)} = -\dot{u}\dot{u}, \quad\mbox{and}\quad (vv{\dot)} = -\dot{v}\dot{v}.
 \end{equation}
We noticed a similar sign change in Equation~(\ref{max22}).

\par

In order to construct the $z$ component in this
$O(3)$ space, let us first
consider:
\begin{equation}\label{max51}
f_{z} = \frac{1}{2}\left[(uv + vu) -
 \left(\dot{u}\dot{v} + \dot{v}\dot{u}\right)\right] ,
\qquad
g_{z} = \frac{1}{2i}\left[(uv + vu) +
\left(\dot{u}\dot{v} + \dot{v}\dot{u}\right)\right] .
\end{equation}

Here, $f_{z}$ and $g_{z}$ are respectively symmetric and anti-symmetric
under the dot conjugation or the parity operation. These quantities
are invariant under the boost along the $z$ direction. They are also
invariant under rotations around this axis, but they are not invariant
under boosts along or rotations around the $x$ or $y$ axis. They are
different from the scalars given in Equation~(\ref{max11}).
\par
Next, in order to construct the $x$ and $y$ components, we start with
$f_{\pm}$ and $g_{\pm}$ as:
\begin{eqnarray}\label{max53}
&{}& f_{+} = \frac{1}{\sqrt{2}}\left(uu - \dot{u}\dot{u}\right),
 \qquad f_{-} = \frac{1}{\sqrt{2}}\left(vv - \dot{v}\dot{v}\right),
 \nonumber\\[1ex]
&{}& g_{+} = \frac{1}{\sqrt{2}i}\left(uu + \dot{u}\dot{u}\right),
 \qquad
 g_{-} = \frac{1}{\sqrt{2}i}\left(vv + \dot{v}\dot{v}\right).
\end{eqnarray}

Then:
\begin{eqnarray}\label{max55}
&{}& f_{x} = \frac{1}{\sqrt{2}}\left(f_{+} + f_{-}\right)
= \frac{1}{2}\left[ \left(uu + vv\right) -
 \left(\dot{u}\dot{u} + \dot{v}\dot{v}\right)\right], \nonumber\\[1ex]
&{}& f_{y} = \frac{1}{\sqrt{2}i}\left(f_{+} - f_{-}\right)
= \frac{1}{2i}\left[ \left(uu - vv\right)
- \left(\dot{u}\dot{u} - \dot{v}\dot{v}\right)\right] ,
\end{eqnarray}
and:
\begin{eqnarray}\label{max57}
&{}& g_{x} = \frac{1}{\sqrt{2}}\left(g_{+} + g_{-}\right)
= \frac{1}{2}\left[ \left(uu + vv\right) +
 \left(\dot{u}\dot{u} + \dot{v}\dot{v}\right)\right], \nonumber\\[1ex]
&{}& g_{y} = \frac{1}{\sqrt{2}i}\left(g_{+} - g_{-}\right)
= \frac{1}{2i}\left[ \left(uu - vv\right)
+ \left(\dot{u}\dot{u} - \dot{v}\dot{v}\right)\right] .
\end{eqnarray}

Here, $f_{x}$ and $f_{y}$ are symmetric under dot conjugation, while
$g_{x}$ and $g_{y}$ are anti-symmetric.

\par
Furthermore, $f_{z}, f_{x}$ and $f_{y}$ of Equations~(\ref{max51}) and
(\ref{max55}) transform like a three-dimensional vector. The~same can be
said for $g_{i}$ of Equations~(\ref{max51}) and (\ref{max57}). Thus, they can be
grouped into the second-rank tensor:
\begin{equation}\label{max59}
 \pmatrix{ 0 & -f_z & -f_x & -f_y \cr f_z & 0 & -g_y & g_x \cr
 f_x & g_y & 0 & -g_z \cr f_y & -g_x & g_z & 0
 } ,
\end{equation}
whose Lorentz-transformation properties are well known. The $g_{i}$
components change their signs under space inversion, while the $f_{i}$
components remain invariant. They are like the electric and magnetic fields,
respectively.

\par

If the system is Lorentz-boosted, $f_{i}$ and $g_{i}$ can
be computed from Table~\ref{tab77}. We are now interested in the symmetry of
photons by taking the massless limit. Thus, we keep only the terms that
become larger for larger values of $\eta$. Thus,
\begin{eqnarray}
&{}& f_{x} \rightarrow \frac{1}{2} \left(uu - \dot{v}\dot{v}\right), \qquad
 f_{y} \rightarrow \frac{1}{2i} \left(uu + \dot{v}\dot{v}\right),
 \nonumber\\[2ex]
&{}& g_{x} \rightarrow \frac{1}{2i} \left(uu + \dot{v}\dot{v}\right),
 \qquad
 g_{y} \rightarrow -\frac{1}{2} \left(uu - \dot{v}\dot{v}\right) ,
\end{eqnarray}
in the massless limit.

\par

Then, the tensor of Equation~(\ref{max59}) becomes:
\begin{equation}\label{max52}
\pmatrix{ 0 & 0 & -E_{x} & -E_{y} \cr
0 & 0 & -B_{y} & B_{x} \cr
E_{x} & B_{y} & 0 & 0 \cr
E_{y} & -B_{x} & 0 & 0 } ,
\end{equation}
with:
\begin{eqnarray}\label{max60}
&{}& E_{x} \simeq \frac{1}{2} \left(uu - \dot{v}\dot{v}\right),
 \qquad
 E_{y} \simeq \frac{1}{2i} \left(uu + \dot{v}\dot{v}\right),
 \nonumber\\[2ex]
&{}& B_{x} = \frac{1}{2i} \left(uu + \dot{v}\dot{v}\right),
 \qquad
 B_{y} = -\frac{1}{2} \left(uu - \dot{v}\dot{v}\right) .
\end{eqnarray}

\par

The electric and magnetic field components are perpendicular to
each other. Furthermore,
\begin{equation}
B_{x} = E_{y}, \quad\mbox\quad B_{y} = - E_{x} .
\end{equation}

\par

In order to address symmetry of photons, let us go back to Equation~(\ref{max53}).
In the massless limit,
\begin{equation}\label{max63}
B_{+} \simeq E_{+} \simeq uu, \qquad B_{-} \simeq E_{-}
 \simeq \dot{v}\dot{v}.
\end{equation}

The gauge transformations applicable to $u$
and $\dot{v}$ are
the two-by-two matrices:
\begin{equation}\label{3tri09}
\pmatrix{1 & -\gamma \cr 0 & 1}, \quad\mbox{and}\quad
\pmatrix{1 & 0 \cr \gamma & 1} ,
\end{equation}
respectively.
Both $u$ and $\dot{v}$ are invariant under gauge transformations,
while $\dot{u}$ and $v$ are not.
\par
The $B_{+}$ and $E_{+}$ are for the photon spin along the $z$ direction,
while $B_{-}$ and $E_{-}$ are for the opposite~direction.

\subsection{Higher Spins}

Since Wigner's original book of 1931~\cite{wig31,wig59}, the rotation
group, without Lorentz transformations, has been extensively discussed in the
literature~\cite{gelfand63,condon51,hamer62}. One of the main issues was
how to construct the most general spin state from the two-component
spinors for the spin-1/2 particle.

Since there are two states for the spin-1/2 particle, four states can be
constructed from two spinors, leading to one state for the spin-0 state and three
spin-1 states. With three spinors, it is possible to construct four spin-3/2
states and two spin-1/2 states, resulting in six states. This partition process
is much more complicated~\cite{fkr71,hkn80ajp} for the case of three spinors.
Yet, this partition process is possible for all higher spin states.

\begin{figure}[thb]
\centerline{\includegraphics[scale=4.0]{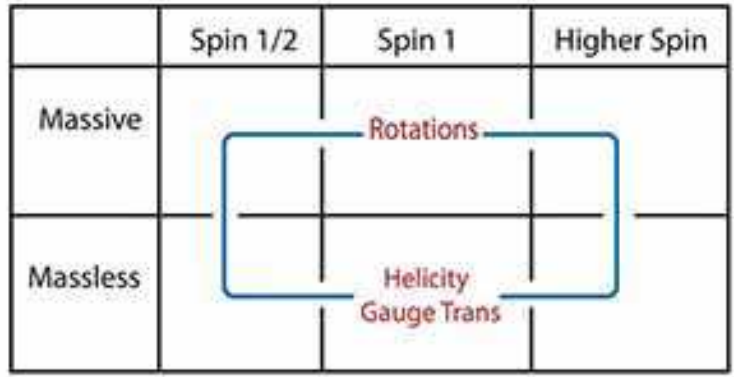}}
\caption{Unified picture of massive and massless particles. The gauge
transformation is a Lorentz-boosted rotation matrix and is applicable
to all massless particles. It is possible to construct higher-spin
states starting from the four states of the spin-1/2 particle in
the Lorentz-covariant world. }\label{gauge33}
\end{figure}

In the Lorentz-covariant world, there are four states for each spin-1/2 particle.
With two spinors, we end up with sixteen (4 $\times$ 4) states, and they are
tabulated in Table~\ref{tab77}. There should be 64 states for three spinors
and 256 states for four spinors.
We now know how to Lorentz-boost those spinors. We also know that the
transverse rotations become gauge transformations in the limit of zero-mass
or infinite-$\eta$. It is thus possible to bundle all of them into the table
given in Figure~\ref{gauge33}.
%----------------------------------------------------------------------
%----------------------------------------------------------------------

\par
In the relativistic regime, we are interested in photons and gravitons.
As was noted in Sections~\ref{4vec} and~\ref{tensor}, the observable
components are invariant under gauge transformations. They~are also the
 terms that become largest for large values of $\eta$.

\par
We have seen in Section~\ref{tensor} that the photon state consists of $uu$ and
$\dot{v}\dot{v}$ for those whose spins are parallel and anti-parallel to
the momentum, respectively. Thus, for spin-2 gravitons, the states must be
$uuuu$ and $\dot{v}\dot{v}\dot{v}\dot{v}$, respectively.

\par
In his effort to understand photons and gravitons, Weinberg constructed his
states for massless particles~\cite{wein64b}, especially photons and
gravitons~\cite{wein64c}. He started with the conditions:
\begin{equation}
N_{1}|\mbox{state}> = 0, \quad\mbox{and}\quad N_{2}|\mbox{state}> = 0,
\end{equation}
where $N_{1}$ and $N_{2}$ are defined in Equation (\ref{n12}). Since they are
now known as the generators of gauge transformations, Weinberg's states are
gauge-invariant states. Thus, $uu$ and $\dot{v}\dot{v}$ are Weinberg's
states for photons, and $uuuu$ are $\dot{v}\dot{v}\dot{v}\dot{v}$ are
Weinberg's states for gravitons.

\section{Concluding Remarks}

Since the publication of Wigner's original paper~\cite{wig39}, there have
been many papers written on the subject. The issue is how to construct
subgroups of the Lorentz group whose transformations do not change the
momentum of a given particle. The traditional approach to this problem
has been to work with a fixed mass, which remains invariant under Lorentz
transformation.

In this paper, we have presented a different approach. Since, we are
interested in transformations that leave the momentum invariant, we do not
change the momentum throughout mathematical processes. Figure~\ref{alphachi}
tells the difference. In our approach, we fix the momentum, and we allow
transitions from one hyperbola to another analytically with one transformation
matrix. It is an interesting future problem to see what larger group
can accommodate this process.

Since the purpose of this paper is to provide a simpler mathematics for
understanding the physics of Wigner's little groups, we used the two-by-two
$SL(2,c)$ representation, instead of four-by-four matrices, for the Lorentz
group throughout the paper. During this process, it was noted in
Section~\ref{dirac} that the Dirac equation is a representation of Wigner's
little group.

We also discussed how to construct higher-spin states starting from
four-component spinors for the spin-1/2 particle. We studied how the
spins can be added in the Lorentz-covariant world, as illustrated in
Figure~\ref{gauge33}.

\end{document}